\documentclass[runningheads]{llncs}
\usepackage{amsmath,amssymb}
\usepackage{semantic}
\usepackage{stmaryrd}
\usepackage{mathtools}
\usepackage{bbm}
\usepackage{tabularx}
\usepackage{listings}
\usepackage[dvipsnames]{xcolor}
\usepackage{xcolor}
\usepackage{amsmath}
\usepackage{amsfonts}
\usepackage{amssymb}
\usepackage{fancyref}
\usepackage[lined,linesnumbered,algoruled]{algorithm2e}
\usepackage{algpseudocode}
\usepackage{fontawesome}
\usepackage{todonotes}
\usepackage[bookmarks=true]{hyperref}
\usepackage{tikz}
\usepackage{letltxmacro} 
\usepackage{graphicx}
\usepackage{makeidx} 
\makeindex 
\usepackage{multirow}
\usepackage{parcolumns}
\usepackage{pgfplots}

\usepackage{kg_macros}

\usepackage{biblatex}
\begin{document}
\title{WShEx: A language to describe and validate Wikibase entities}
%
%
\author{Jose Emilio labra Gayo\inst{1}\orcidID{0000-0001-8907-5348}}
\authorrunning{J. E. Labra G.}
%
\institute{WESO Research Group - University of Oviedo, Spain \\
\email{labra@uniovi.es}\\
\url{http://labra.weso.es}
}
\maketitle              
\begin{abstract}
Wikidata is one of the most successful Semantic Web projects. 
 Its underlying Wikibase data model departs from RDF with the inclusion of several features like 
 qualifiers and references, built-in datatypes, etc.  
Those features are serialized to RDF for content negotiation, RDF dumps and in the SPARQL endpoint.
Wikidata adopted the entity schemas namespace using the ShEx language to describe and validate the RDF serialization of Wikidata entities. 
In this paper we propose WShEx, a language inspired by ShEx that directly supports the Wikibase data model and can be used
 to describe and validate Wikibase entities. 
 The paper presents a the abstract syntax and semantic of the WShEx language.

\end{abstract}
\newcommand\vartextvisiblespace[1][.5em]{%
  \makebox[#1]{%
    \kern.07em
    \vrule height.3ex
    \hrulefill
    \vrule height.3ex
    \kern.07em
  }
}

\newcommand{\mi}[1]{\ensuremath{\mathit{#1}}}
\newcommand{\cM}[1]{\ensuremath{\text{#1}}}

\newcommand\myfont[1]{\ensuremath{\mathcal{#1}}}


\newcommand\ShapeLabel{L}

\newcommand\triple[3]{\ensuremath{\langle #1,#2,#3 \rangle}}
\newcommand\quadruple[4]{\ensuremath{\langle #1,#2,#3,#4\rangle}}
\newcommand\Graph{\myfont{G}}
\newcommand{\tcs}[2]{tcs(#1,#2)}

\newcommand\VertSet{\myfont{V}}
\newcommand\NodeSet{\myfont{N}}
\newcommand\EdgeSet{\myfont{E}}
\newcommand\LabelSet{\myfont{L}}
\newcommand\PLabelSet{\myfont{T}}
\newcommand\MsgSet{\myfont{M}}

\newcommand{\ItemSet}{\myfont{Q}}
\newcommand{\PropSet}{\myfont{P}}
\newcommand{\EntitySet}{\myfont{E}}
\newcommand{\ValueSet}{\myfont{V}}
\newcommand{\DataValueSet}{\myfont{D}}
\newcommand{\StmtSet}{\rho}
\newcommand{\SubjectSet}{\myfont{S}}
\newcommand{\ObjectSet}{\myfont{O}}
\newcommand{\Nat}{\mathbb{N}}

\newcommand{\qs}[1]{\ensuremath{#1}}
\newcommand{\condType}[1]{\ensuremath{hasType_{#1}}}

\newcommand{\neighsGraph}[3]{#1^{#2}_{#3}}
\newcommand{\partition}[1]{part(#1)}
\newcommand{\FinSet}[1]{\ensuremath{FinSet(#1)}}

\newcommand{\emptyGraph}{\ensuremath{\emptyset}}
\newcommand{\EmptyGraph}{\ensuremath{\emptyset}}
\newcommand{\addTriple}{\ensuremath{\rtimes}}
\newcommand{\unionGraphs}{\ensuremath{\cup}}
\newcommand\neighs[2]{\ensuremath{neighs(#1,#2)}}
\newcommand\nodes[1]{\ensuremath{nodes(#1)}}
\newcommand{\node}{\mi{n}}
\newcommand{\lbl}{\mi{l}}

\newcommand{\typing}{\tau}
\newcommand{\emptyTyping}{[]}
\newcommand{\hasType}[2]{\ensuremath{#1@#2}}
\newcommand{\hasNoType}[2]{\ensuremath{#1@!#2}}
\newcommand\EmptyTyping{[]}
\newcommand\addType[2]{\ensuremath{#1@#2}}
\newcommand\AddTyping[3]{\ensuremath{#1@#2 : #3}}
\newcommand\RemoveTyping[3]{\ensuremath{#1@!#2 : (#3 \setminus #1@#2)}}
\newcommand\Combine[2]{\ensuremath{#1\uplus#2}}
\newcommand\CombineOr[2]{\ensuremath{#1\parallel#2}}
\newcommand{\HasType}[3]{\ensuremath{#1@#2\in{}#3}}
\newcommand{\HasNoType}[3]{\ensuremath{#1@!#2\in{}#3}}
\newcommand\FailTyping{\ensuremath{\mathbbm{E}}}

\newcommand\arc[2]{\ensuremath{\vartextvisiblespace\xrightarrow{#1}#2}}
\newcommand{\qualified}[4] {\ensuremath{\arc{#1}{#2}\{#3,#4\}}}
\newcommand{\qualifiedcard}[3] {\ensuremath{\arc{#1}{#2}~#3}}

\newcommand{\semantics}[4] {\ensuremath{|[#1|]^{#2,#3,#4}}}
\newcommand{\nodeSelector}[2]{\ensuremath{|[#1|]^{#2}}}
\newcommand{\shape}{\ensuremath{\phi}}
\newcommand{\shapeRef}[1] {\ensuremath{@#1}}
\newcommand{\shapeTrue}{\ensuremath{\top}}
\newcommand{\sAnd}{\ensuremath{\wedge}}
\newcommand{\sNot}{\ensuremath{\lnot}}
\newcommand{\unbounded}{\ensuremath{*}}

\newcommand{\shapesSet}{\operatorname{\mathit{shapes}}}

\newcommand\ShapeN[1]{\ensuremath{\ShapeGen{#1}{n}}}
\newcommand\ShapeO[1]{\ensuremath{\ShapeGen{#1}{o}}}
\newcommand\ShapeGen[2]{\ensuremath{\Triples{}^{#1}_#2}}

\newcommand{\ShapeNStar}{\ensuremath{\Triples^{*}}}
\newcommand{\SE}{\ensuremath{\mathcal{S}_n(E)}}
\newcommand\Shape[1]{\ensuremath{\mathcal{S}_n|[#1|]}}
\newcommand{\Triples}{\ensuremath{\Sigma}}

\newcommand\deriv[2]{\ensuremath{\partial_{#1}(#2)}}
\newcommand\nullable{\ensuremath{\nu}}

\newcommand{\defineSchema}{\ensuremath{\longmapsto}}

\newcommand{\oom}{\ensuremath{\infty}}

\newcommand{\Vs}{\ensuremath{V_s}}
\newcommand{\Vp}{\ensuremath{V_p}}
\newcommand{\Vo}{\ensuremath{V_o}}

\newcommand\SymbolSet{\ensuremath{\Delta}}
\newcommand{\bagw}{\ensuremath{w}}
\newcommand{\emptyBag}{\ensuremath{\epsilon}}
\newcommand{\bag[1]}{\{|#1|\}}
\newcommand{\bagSem}[1]{\ensuremath{|[#1|]}}
\newcommand{\matchRbeOpen}[2]{#1\approx{}#2}
\newcommand{\matchRbeClosed}[2]{#1\approxeq{}#2}
\newcommand{\matchRbe}[2]{#1\approxeq{}#2}

\newcommand{\eachOf}[2]{\ensuremath{#1;#2}}
\newcommand{\oneOf}[2]{\ensuremath{#1\mid{}#2}}
\newcommand{\someOf}[2]{\ensuremath{#1\mid{}#2}}

\newcommand{\tripleConstraint}[4]{\ensuremath{\qualified{#1}{#2}{#3}{#4}}}
\newcommand{\shapes}[1]{\ensuremath{shapes^{#1}}}
\newcommand{\props}[1]{\ensuremath{preds(#1)}} 
\newcommand{\preds}[1]{\ensuremath{preds(#1)}}
\newcommand{\descendants}[1]{\ensuremath{descendants(#1)}}
\newcommand{\descendantsS}[2]{\ensuremath{descendants_{#1}(#2)}}

\newcommand\Pass{\ensuremath{\mathbbm{p}}}
\newcommand\Fail{\ensuremath{\mathbbm{f}}}
\newcommand\code{\textit}

\newcommand{\Dunno}{\ensuremath{\theta}}
\newcommand{\Empty}{\ensuremath{\varnothing}}
\newcommand{\None}{\ensuremath{\varnothing}}
\newcommand\Over{\ensuremath{\mathbbm{e}}}
\newcommand\Open{\ensuremath{\mathbbm{z}}}

\newcommand{\eq}{==}

\newcommand{\emptyRule}{\ensuremath{\varepsilon}}
\newcommand{\matchShape}{\ensuremath{\simeq_s}}
\newcommand{\matchRule}{\ensuremath{\simeq_r}}
\newcommand{\matchName}{\ensuremath{\simeq_n}}
\newcommand{\matchValue}{\ensuremath{\simeq_v}}

\newcommand{\match}{\ensuremath{\simeq}}
\newcommand{\context}{\ensuremath{\Gamma}}
\newcommand{\emptyEr}{\ensuremath{\varepsilon}}
\newcommand{\andEr}{\ensuremath{\parallel}}
\newcommand{\orEr}{\ensuremath{|}}
\newcommand{\fail}{\ensuremath{\emptyset}}

\newcommand{\iriKind}{\ensuremath{\cM{IRI}}}

\newcommand{\Schema}{\ensuremath{\myfont{S}}}
\newcommand{\schema}{\Schema}

\newcommand{\AbstractSet}{\ensuremath{\myfont{A}}}
\newcommand{\schemaDef}{\ensuremath{\delta}}
\newcommand{\ShapeSet}{\ensuremath{\myfont{S}}}
\newcommand\IRISet{\myfont{I}}
\newcommand\BNodeSet{\myfont{B}}
\newcommand\LitSet{\mathit{Lit}}

\newcommand{\bnodeKind}{\ensuremath{\cM{BNode}}}
\newcommand{\datatype}[1]{\ensuremath{datatype(#1)}}
\newcommand{\bcond}[1]{\ensuremath{\Psi_{#1}}}
\newcommand{\predSpec}{\ensuremath{p}}

\newcommand{\openQs}[1]{\ensuremath{\lfloor#1\rfloor}}
\newcommand{\closeQs}[1]{\ensuremath{\lceil#1\rceil}}
\newcommand{\eachOfQs}[2]{#1\,,\,#2}
\newcommand{\oneOfQs}[2]{\ensuremath{#1\mid{}#2}}

\newcommand{\false}{\ensuremath{0}}
\newcommand{\iri}[1]{{\color{blue}\textit{:#1}}}
\newcommand{\iriq}[2]{{\color{red}#1}{\color{blue}\textit{:#2}}}
\newcommand{\literal}[1]{\ensuremath{\text{\texttt{#1}}}}
\newcommand{\extends}[2]{\ensuremath{extends\,@#1\,#2}}
\newcommand{\restricts}[2]{\ensuremath{restricts\,@#1\,#2}}
\newcommand{\trans}[2]{\ensuremath{\langle#2\rangle^{#1}}}
\newcommand{\transform}[2]{\trans{#1}{\iri{#2}}}
\newcommand{\embed}[3]{\ensuremath{\langle#2\rtimes#3\rangle^{#1}}}

\makeatletter
\newcount\my@repeat@count
\newcommand{\myrepeat}[2]{%
  \begingroup
  \my@repeat@count=\z@
  \@whilenum\my@repeat@count<#1\do{#2\advance\my@repeat@count\@ne}%
  \endgroup
}
\makeatother
\newcommand{\sep}[1]{\myrepeat{#1}{\space}}

\newcommand*{\SavedLstInline}{}
\LetLtxMacro\SavedLstInline\lstinline
\DeclareRobustCommand*{\lstinline}{%
  \ifmmode
    \let\SavedBGroup\bgroup
    \def\bgroup{%
      \let\bgroup\SavedBGroup
      \hbox\bgroup
    }%
  \fi
  \SavedLstInline
}

\lstdefinestyle{inlinecode}{basicstyle=\ttfamily\footnotesize\bfseries}
\renewcommand\c{\lstinline[style=inlinecode]}

\SetKwProg{Def}{def}{$\,$:}{}
\SetKwProg{Defn}{def}{~$=$}{}
\SetKw{defn}{def}
\newcommand{\DefInline}[2]{\defn #1 = #2}
\SetKwProg{DefnCustom}{\defn}{}{}
\SetKw{Let}{let}
\SetKwInput{KwIn}{Input}
\SetKwFor{ForEach}{foreach}{}{}
\SetKw{Or}{or}
\SetKw{And}{and}
\SetKwIF{If}{ElseIf}{Else}{if}{then}{else if}{else}{endif}
\SetKw{Match}{match}
\SetKw{MyIf}{if}
\SetKw{MyThen}{then}
\SetKw{MyElse}{else}
\SetKw{Case}{case}
\SetKwBlock{Let}{let}{in}
\SetKw{In}{in}
\SetKw{MapTo}{\ensuremath{\;\;\Rightarrow\;\;}}
\SetKwBlock{Block}{}{}
\newcommand{\assign}{\ensuremath{~\mathtt{:=}~}}
\newcommand{\algocomment}[1]{\text{//$\,$#1}}

\newcommand{\blockskip}{\smallskip}
\newcommand{\done}{\ensuremath{\mathit{done}}}

\def\algorithmsize{small}
\def\algorithmheadersize{\algorithmsize}
\renewcommand\AlCapFnt{\normalfont\bfseries\small}
\setlength{\textfloatsep}{1.0ex} 
\setlength{\floatsep}{1.0ex} 


\newcommand{\mycomment}[1]{\emph{// #1}}
\newcommand{\shapesGraph}{\ensuremath{S_g}}

\newcommand{\invPP}[1]{\ensuremath{~\hat{}#1}}
\newcommand{\seqPP}[2]{\ensuremath{#1\cdot{}#2}}
\newcommand{\altPP}[2]{\ensuremath{#1\lor{}#2}}
\newcommand{\notPP}[1]{\ensuremath{!#1}}
\newcommand{\starPP}[1]{\ensuremath{#1^{*}}}
\newcommand{\semPP}[2]{\ensuremath{|[#1|]^{#2}}}
\newcommand{\countProp}[6]{\ensuremath{\#_{#1,#6}^{#2,#3,#4,#5}}}
\newcommand{\countAll}[3]{\ensuremath{\#_{#1}^{#2,#3}}}
\newcommand{\conforms}[4]{#1,#2,#3\vDash#4}
\newcommand{\conformsNot}[4]{#1,#2,#3\nvDash#4}
\newcommand{\conformsTE}[4]{#1,#2,#3\Vdash#4}
\newcommand{\conformsQs}[4]{#1,#2,#3\vdash#4}

\newcommand{\slurpSE}[5]{#1,#2,#3\vDash#4\rightsquigarrow#5}
\newcommand{\slurpTE}[5]{#1,#2,#3\Vdash#4\rightsquigarrow#5}
\newcommand{\slurpQs}[5]{#1,#2,#3\vdash#4\rightsquigarrow#5}

\definecolor{forestgreen}{rgb}{34,139,34}
\definecolor{orangered}{rgb}{239,134,64}
\definecolor{darkblue}{rgb}{0.0,0.0,0.6}
\definecolor{gray}{rgb}{0.4,0.4,0.4}

\definecolor{LightGray}{rgb}{0.97,0.97,0.97}
\lstdefinelanguage{SPARQL}{
  basicstyle=\small\ttfamily,
  backgroundcolor=\color{LightGray},
  columns=fullflexible,
  breaklines=false,
  sensitive=true,
  frame=bt,
  aboveskip=1em,
  belowskip=1em,
  xleftmargin=.5em,
  xrightmargin=.5em,
  framexleftmargin=.5em,
  framextopmargin=.5em,
  framexbottommargin=.5em,
  framexrightmargin=.5em,
  tabsize = 2,
  showstringspaces=false,
  morecomment=[l][\color{gray}]{\#},       
  morecomment=[n][\color{blue}]{<http}{>}, 
  morestring=[b][\color{OliveGreen}]{\"},  
  keywordsprefix=?,
  classoffset=0,
  keywordstyle=\color{Sepia},
  morekeywords={},
  classoffset=1,
  keywordstyle=\color{Purple},
  morekeywords={rdf,rdfs,owl,xsd,purl},
  classoffset=2,
  keywordstyle=\color{MidnightBlue},
  morekeywords={
    SELECT,CONSTRUCT,DESCRIBE,ASK,WHERE,FROM,NAMED,PREFIX,BASE,OPTIONAL,
    FILTER,GRAPH,LIMIT,OFFSET,SERVICE,UNION,EXISTS,NOT,BINDINGS,MINUS,a
  }
}

\lstdefinestyle{Cypher}{
 numberblanklines=false, 
 morekeywords={MATCH, CREATE, RETURN, WITH, ORDER, BY, WHERE, FOREACH, DELETE, OPTIONAL, UNWIND, SKIP, LIMIT, SET, REMOVE, MERGE, CALL, UNION, USE, LOAD
 },
 sensitive=false
}


\lstdefinestyle{SHACL}{
    numberblanklines=false, 
    keywords={prefix, @prefix, 
   sh:and, sh:class, sh:closed, sh:constraintComponent, sh:conforms,
   sh:datatype, sh:disjoint, sh:equals, 
   sh:flags, sh:focusNode, 
   sh:hasValue,
   sh:ignoredProperties, sh:in, sh:js, sh:jsLibrary, sh:jsLibraryURL, sh:jsFunctionName, sh:JSConstraint,
   sh:languageIn, sh:lessThan, sh:lessThanOrEquals, 
   sh:maxCount, sh:maxExclusive, sh:maxInclusive, sh:maxLength, sh:minCount, sh:minExclusive, sh:minInclusive, sh:minLength, 
   sh:node, sh:nodeKind, sh:not, 
   sh:optional, sh:or, 
   sh:path, sh:pattern, sh:property, sh:qualifiedMaxCount, sh:qualifiedMinCount, sh:qualifiedValueShape,
   sh:qualifiedValueShape, sh:qualifiedValueShapesDisjoint, sh:sparql, 
   sh:resultSeverity, sh:resultPath, sh:resultMessage, 
   sh:sourceConstraintComponent, sh:sourceShape,
   sh:targetClass, sh:targetNode, sh:targetObjectsOf, sh:targetSubjectsOf,
   sh:uniqueLang, 
   sh:value, 
   sh:xone,
   sh:IRI, sh:Literal, sh:NonLiteral, sh:BlankNodeOrLiteral, sh:BlankNodeOrIRI, sh:IRIOrLiteral,
   sh:NodeShape, sh:PropertyShape, sh:ValidationReport, sh:ValidationResult, sh:Violation
   },
  morekeywords={
   cex:,cdt:,dc:,dct:,dbo:,dbr:,doap:,ex:,foaf:,lemon:,org:,owl:,qb:,qp:,rdf:,rdfs:,schema:,skos:,skosxl:,void:,wr:,wt:,wf:,xsd:,sh:,sx:
  }
  alsoletter={:,@},
  comment=[l]{\#},
  morestring=[b]',
  morestring=[b]",
  alsoletter={:}
}


\newcommand\JSONnumbervaluestyle{\color{blue}}
\newcommand\JSONstringvaluestyle{\color{red}}

\newif\ifcolonfoundonthisline

\makeatletter

\lstdefinestyle{json}
{
  showstringspaces    = false,
  keywords            = {false,true},
  alsoletter          = 0123456789.,
  morestring          = [s]{"}{"},
  stringstyle         = \ifcolonfoundonthisline\JSONstringvaluestyle\fi,
  MoreSelectCharTable =%
    \lst@DefSaveDef{`:}\colon@json{\processColon@json},
  basicstyle          = \ttfamily,
  keywordstyle        = \ttfamily\bfseries,
}

\newcommand\processColon@json{%
  \colon@json%
  \ifnum\lst@mode=\lst@Pmode%
    \global\colonfoundonthislinetrue%
  \fi
}

\lst@AddToHook{Output}{%
  \ifcolonfoundonthisline%
    \ifnum\lst@mode=\lst@Pmode%
      \def\lst@thestyle{\JSONnumbervaluestyle}%
    \fi
  \fi
 \lsthk@DetectKeywords%
}

\lst@AddToHook{EOL}%
  {\global\colonfoundonthislinefalse}

\makeatother

\definecolor{LightGray}{rgb}{0.97,0.97,0.97}

\lstdefinelanguage{Turtle}{
  basicstyle=\small\ttfamily,
  backgroundcolor=\color{LightGray},
  columns=fullflexible,
  breaklines=false,
  sensitive=true,
  frame=bt,
  aboveskip=1em,
  belowskip=1em,
  xleftmargin=.5em,
  xrightmargin=.5em,
  framexleftmargin=.5em,
  framextopmargin=.5em,
  framexbottommargin=.5em,
  framexrightmargin=.5em,
  tabsize = 2,
  showstringspaces=false,
  morecomment=[l][\color{gray}]{\#},   
  morecomment=[n][\color{blue}]{<}{>}, 
  morestring=[b][\color{OliveGreen}]{\"},  
  classoffset=0,
  keywordstyle=\color{Sepia},
  morekeywords={},
  classoffset=1,
  keywordstyle=\color{Purple},
  morekeywords={rdf,rdfs,owl,xsd,purl,wdt,wd,p,ps,pq, wikibase},
  classoffset=2,
  keywordstyle=\color{MidnightBlue},
  morekeywords={
   a
  }
}

\lstdefinelanguage{ShExC}{
  basicstyle=\small\ttfamily,
  backgroundcolor=\color{LightGray},
  columns=fullflexible,
  breaklines=false,
  sensitive=true,
  frame=bt,
  aboveskip=1em,
  belowskip=1em,
  xleftmargin=.5em,
  xrightmargin=.5em,
  framexleftmargin=.5em,
  framextopmargin=.5em,
  framexbottommargin=.5em,
  framexrightmargin=.5em,
  tabsize = 2,
  showstringspaces=false,
  morecomment=[l][\color{gray}]{\#},   
  morecomment=[n][\color{blue}]{<}{>}, 
  morestring=[b][\color{OliveGreen}]{\"},  
  classoffset=0,
  keywordstyle=\color{Sepia},
  morekeywords={},
  classoffset=1,
  keywordstyle=\color{Purple},
  morekeywords={rdf,rdfs,owl,xsd,purl,wdt,wd,p,ps,pq},
  classoffset=2,
  keywordstyle=\color{MidnightBlue},
  morekeywords={
   CLOSED, EXTRA, ABSTRACT, EXTENDS,
   IRI,BNODE,MININCLUSIVE,MAXINCLUSIVE,MINEXCLUSIVE,MAXINCLUSIVE
  }
}

\lstdefinelanguage{WShEx}{
  basicstyle=\small\ttfamily,
  backgroundcolor=\color{LightGray},
  columns=fullflexible,
  breaklines=false,
  sensitive=true,
  frame=bt,
  aboveskip=1em,
  belowskip=1em,
  xleftmargin=.5em,
  xrightmargin=.5em,
  framexleftmargin=.5em,
  framextopmargin=.5em,
  framexbottommargin=.5em,
  framexrightmargin=.5em,
  tabsize = 2,
  showstringspaces=false,
  morecomment=[l][\color{gray}]{\#},   
  morecomment=[n][\color{blue}]{<}{>}, 
  morestring=[b][\color{OliveGreen}]{\"},  
  classoffset=0,
  keywordstyle=\color{Sepia},
  morekeywords={},
  classoffset=1,
  keywordstyle=\color{Purple},
  morekeywords={rdf,rdfs,owl,xsd,purl,wdt,wd,p,ps,pq},
  classoffset=2,
  keywordstyle=\color{MidnightBlue},
  morekeywords={
   CommonsMedia, GlobeCoordinate, Item, Property, String,
   MonolingualText, ExternalIdentifier, Quantity, 
   Time, URL, MathematicalExpression, GeographicShape, MusicalNotation, TabularData, Lexeme, Form, Sense, 
   CLOSED, EXTRA
  }
}

\lstdefinestyle{SHACLC}{
  keywords={closed,ignoredProperties,true,false,IRI,BlankNode,Literal,BlankNodeOrLiteral,IRIOrLiteral
  , prefix,
   cex:,cdt:,dc:,dct:,dbo:,dbr:,doap:,ex:,foaf:,lemon:,org:,owl:,qb:,qp:,rdf:,rdfs:,sh:,sx:,schema:,skos:,skosxl:,void:,wr:,wt:,wf:,xsd:
},
 ndkeywordstyle=\color{darkgray}\bfseries,
  identifierstyle=\color{black},
  sensitive=false,
  commentstyle=\color{purple}\ttfamily,
  stringstyle=\color{brown}\ttfamily,
  morestring=[b]',
  morestring=[b]",
  alsoletter={:}
}

\lstdefinestyle{SQL}{numberblanklines=true, 
    morekeywords={CREATE,TABLE,ENUM,FOREIGN,KEY,ENUM,REFERENCES}}

\lstdefinestyle{HTML} {
    language=HTML,
    extendedchars=true, 
    breaklines=true,
    breakatwhitespace=true,
    emph={},
    emphstyle=\color{red},
    basicstyle=\ttfamily,
    columns=fullflexible,
    commentstyle=\color{gray}\upshape,
    morestring=[b]",
    morecomment=[s]{<?}{?>},
    morecomment=[s][\color{forestgreen}]{<!--}{-->},
    keywordstyle=\color{orangered},
    stringstyle=\ttfamily\color{orangered}\normalfont,
    tagstyle=\color{darkblue},
    morekeywords={attribute,xmlns,version,type,release},
    otherkeywords={attribute=, xmlns=},
}

\lstdefinestyle{XML} {
    language=XML,
    extendedchars=true, 
    breaklines=true,
    breakatwhitespace=true,
    emph={},
    emphstyle=\color{red},
    basicstyle=\ttfamily,
    columns=fullflexible,
    commentstyle=\color{gray}\upshape,
    morestring=[b]",
    morecomment=[s]{<?}{?>},
    morecomment=[s][\color{forestgreen}]{<!--}{-->},
    keywordstyle=\color{orangered},
    stringstyle=\ttfamily\color{orangered}\normalfont,
    tagstyle=\color{darkblue},
    morekeywords={attribute,xmlns,version,type,release},
    otherkeywords={attribute=, xmlns=},
}

\lstdefinestyle{RelaxNG} {
    extendedchars=true, 
    breaklines=true,
    breakatwhitespace=true,
    emph={},
    emphstyle=\color{red},
    basicstyle=\ttfamily,
    columns=fullflexible,
    commentstyle=\color{gray}\upshape,
    morestring=[b]",
    morecomment=[s]{<?}{?>},
    morecomment=[s][\color{forestgreen}]{<!--}{-->},
    keywordstyle=\color{darkblue},
    stringstyle=\ttfamily\color{orangered}\normalfont,
    keywords={attribute,element},
    otherkeywords={attribute=, xmlns=},
}

\lstdefinelanguage{JavaScript}{
  keywords={typeof, new, true, false, catch, function, return, null, catch, switch, var, if, in, while, do, else, case, break},
  keywordstyle=\color{blue}\bfseries,
  ndkeywords={class, export, boolean, throw, implements, import, this},
  ndkeywordstyle=\color{darkgray}\bfseries,
  identifierstyle=\color{black},
  sensitive=false,
  comment=[l]{//},
  morecomment=[s]{/*}{*/},
  commentstyle=\color{purple}\ttfamily,
  stringstyle=\color{red}\ttfamily,
  morestring=[b]',
  morestring=[b]"
}

\lstdefinelanguage{scala}{
  morekeywords={abstract,case,catch,class,def,%
    do,else,extends,false,final,finally,%
    for,if,implicit,import,match,mixin,%
    new,null,object,override,package,%
    private,protected,requires,return,sealed,%
    super,this,throw,trait,true,try,%
    type,val,var,while,with,yield},
  otherkeywords={=>,<-,<\%,<:,>:,\#,@},
  sensitive=true,
  morecomment=[l]{//},
  morecomment=[n]{/*}{*/},
  morestring=[b]",
  morestring=[b]',
  morestring=[b]"""
}

\definecolor{mygreen}{rgb}{0,0.6,0}
\definecolor{mygray}{rgb}{0.5,0.5,0.5}
\definecolor{mymauve}{rgb}{0.58,0,0.82}

\lstdefinestyle{inlinecode}{basicstyle=\ttfamily\footnotesize\bfseries}
\renewcommand\c{\lstinline[style=inlinecode]}

\lstset{
    basicstyle=\small\ttfamily,
    captionpos=b,
    frame=single,
    showstringspaces=false,
    escapeinside={\%*}{*)},
    language=ShExC
}


\newcommand{\hrefc}[3][blue]{\href{#2}{\color{#1}{#3}}}%

\newcommand{\elemento}[2]{\ensuremath{\hrefc[violet]{http://www.wikidata.org/entity/#2}{#1}}}
\newcommand{\propiedad}[2]{\ensuremath{\hrefc[darkblue]{http://www.wikidata.org/entity/#2}{#1}}}
\newcommand{\timBl}{\elemento{timBl}{Q80}}
\newcommand{\vintCerf}{\elemento{vintCerf}{Q92743}}
\newcommand{\newHaven}{\elemento{NewHaven}{Q49145}}
\newcommand{\bobDylan}{\elemento{bobDylan}{Q392}}
\newcommand{\London}{\elemento{London}{Q84}}
\newcommand{\CERN}{\elemento{CERN}{Q42944}}
\newcommand{\UK}{\elemento{UK}{Q145}}
\newcommand{\Spain}{\elemento{Spain}{Q29}}
\newcommand{\PA}{\elemento{PA}{Q329157}}
\newcommand{\birthDate}{\propiedad{birthDate}{P569}}
\newcommand{\birthPlace}{\propiedad{birthPlace}{P19}}
\newcommand{\country}{\propiedad{country}{P27}}
\newcommand{\employer}{\propiedad{employer}{P108}}
\newcommand{\awarded}{\propiedad{awarded}{P166}}
\newcommand{\pointTime}{\propiedad{pointTime}{P585}}
\newcommand{\start}{\propiedad{start}{P580}}
\newcommand{\pEnd}{\propiedad{end}{P582}}
\newcommand{\togetherWith}{\propiedad{togetherWith}{P1706}}
\newcommand{\Human}{\elemento{Human}{Q5}}
\newcommand{\instanceOf}{\propiedad{instanceOf}{P31}}

\newcommand{\indexn}[1]{\index{#1}#1}
\newcommand{\indexe}[1]{\index{#1}\emph{#1}}

\newcommand{\Ok}{\ensuremath{\c|Ok|}}
\newcommand{\Failed}{\ensuremath{\c|Failed|}}
\newcommand{\WaitingFor}[3]{\ensuremath{\c|WaitingFor|(#1,#2,#3)}}
\newcommand{\Pending}{\ensuremath{\c|Pending|}}
\newcommand{\PendingLs}{\ensuremath{\c|Pending|(ls)}}
\newcommand{\Undefined}{\ensuremath{\c|Undefined|}}

\newcommand{\Validate}{\ensuremath{\c|Validate|}}
\newcommand{\Checked}[2]{\ensuremath{\c|Checked|(#1,#2)}}
\newcommand{\WaitFor}[1]{\ensuremath{\c|WaitFor|(#1)}}
\newcommand{\status}[2]{\ensuremath{#1(#2)}}
\newcommand{\msgSent}[3]{\ensuremath{#1,#2\rightsquigarrow{}#3}}
\newcommand{\checkLocal}[2]{\ensuremath{\c|checkLocal|(#1,#2)}}
\newcommand{\checkLocalOpen}[2]{\ensuremath{\c|checkLocalOpen|(#1,#2)}}

\newcommand{\checkNeighs}[3]{\ensuremath{\c|checkNeighs|(#1,#2,#3)}}
\newcommand{\fracEmpty}[2]{\genfrac{}{}{0pt}{0}{#1}{#2}}
\newcommand{\vProg}{\c|vProg|}
\newcommand{\tripleConstraints}[1]{\c|tripleConstraints(#1)|}
\newcommand{\rbe}[1]{\ensuremath{\c|rbe|(#1)}}
\newcommand{\combine}[2]{\ensuremath{\c|combine|(#1,#2)}}

\section{Introduction}

\indexn{Wikidata} has become one of the 
 biggest projects which collect human knowledge in the form of linked data according to the Semantic Web view. 
It is collaboratively maintained both by humans and bots, 
which update the contents from external services or databases. 
The set of 
open source tools which run Wikidata is called Wikibase~\footnote{\url{https://wikiba.se/}} and it makes possible 
to create other knowledge graphs following the same data model as Wikidata but with different content and purposes. 
The projects that are using Wikibase are called Wikibase instances 
 and in this paper we will call the style of knowledge graphs obtained using Wikibase instances as Wikibase graphs.

\index{Mediawiki}
\index{MariaDB}
Wikibase was initially created from MediaWiki software which facilitated adoption by the Wikimedia community. 
Internally, Wikidata content is managed by a relational database~\cite{malyshev2018getting}. 
In order to facilitate data analysis and querying, as well as integrating Wikidata within the semantic web ecosystem, 
it adopted a triple store~\footnote{The current technology is Blazegraph\url{https://blazegraph.com/} but there are plans to replace it} whose contents can be retrieved as RDF data according to the linked data principles and exported as an SPARQL endpoint through the Wikidata query service. 

In 2019, wikidata created a new namespace for entity schemas which can be used to describe and validate the RDF serialization of Wikidata entities using ShEx~\cite{Prudhommeaux2014} 
Entity schemas offer a concise language to describe Wikibase entities. 
Users can create new schemas for different purposes and there are ShEx-based tools that can be used to check if entities conform to entity schemas or visualize entity schemas. 
At the time of this writing there are more than 370 entity schemas created\footnote{A directory for entity schemas can be seen at~\url{https://www.wikidata.org/wiki/Wikidata:Database_reports/EntitySchema_directory}} they are still not part of the mainstream workflow employed by wikidata users. 
Although there may be several reasons for this like the lack of better tool support, one aspect that can also affect this situation is that entity schemas describe the RDF serialization of entities, instead of their underlying Wikibase data model. 
This aspect makes entity schemas a bit more verbose and aggravates their usability. 
In this paper, we propose a new language called WShEx which is inspired by ShEx and can be used to directly describe and validate entities based on the Wikibase data model.

The first motivation for the development of WShEx was to create subsets of Wikidata in different domains using a concise and human-readable language. 
In order to process JSON-based Wikidata dumps, it was not possible to directly use entity schemas which describe the RDF serialization, so we developed WShEx, a language similar to ShEx that could be used to describe Wikibase data models directly. 
Some parts of this paper have been extracted from~\cite{Labra2021KGSubsets}, a larger paper where we also describe the different subsetting techniques employed.

%

\section{Wikibase data model}

\index{JSON} \index{RDF}
The Wikibase data model~\footnote{\url{https://www.mediawiki.org/wiki/Wikibase/DataModel}} 
is defined as an abstract data model that can have different serializations like JSON and RDF. 
\index{UML} \index{Wikidata Object Notation}
It is defined using UML data structures and a notation called Wikidata Object Notation. 

\index{Entity} \index{Statement}
Informally, the data model is formed from entities and statements about those entities. 
\index{Item}  \index{Property}
An entity can either be an item or a property. 
An item is usually represented using a \lstinline|Q| followed by a number and can represent any thing like an abstract of concrete concept. 
For example, \href{http://www.wikidata.org/entity/Q80}{Q80} 
represents Tim Berners-Lee in Wikidata.
A property is usually represented by a \lstinline|P| followed by a number and 
represents a relationship between an item and a value. 
For example, \href{http://www.wikidata.org/entity/P19}{P19} represents the property \emph{place of birth} in Wikidata.
The values that can be associated to a property are constrained to belong to some specific datatype. 
There can be compound datatypes like geographical coordinates. 
\index{Datatypes}
Some of Wikibase datatypes are: quantities, 
dates and times, 
geographic locations and shapes, 
monolingual and 
multilingual texts, etc. 

\index{Statement} 
A statement consists of:
\begin{itemize}
	\item A property which is usually denoted using a \lstinline|P| followed by a number.
	\item A declaration about the possible value (in wikibase terms, it is called a \emph{snak}) which can be a specific value, a \lstinline|no value| declaration or a \lstinline|some value| declaration. 
	\item A rank declaration which can be either \lstinline|preferred|, \lstinline|normal| or \lstinline|deprecated|.
	\item Zero or more qualifiers which consist of a list of property-value pairs
	\item Zero or more references which consist of a list of property-value pairs.
\end{itemize}


We define a formal model for Wikibase which is inspired from 
Multi-Attributed Relational Structures (MARS)~\cite{MKT2017}. 
For simplicity, 
we model only qualifiers omitting references, which could be represented in a similar way, 
don't handle the no-value and some-value snaks, 
and represent only one primititve data value~\DataValueSet{}. 

\label{sec:WikibaseDataModel_FormalDefinition}
\begin{definition}[Wikibase graphs] \label{def:wikibaseGraph}
Given a mutually disjoint set of items \ItemSet{}, 
	a set of properties \PropSet{} and 
	a set of data values \DataValueSet{}, 
	a \emph{Wikibase graph} 
	is a tuple $\langle\ItemSet,\PropSet,\DataValueSet,\StmtSet\rangle$ such that
	$\StmtSet\subseteq\EntitySet\times\PropSet\times\ValueSet\times\FinSet{\PropSet\times\ValueSet}$ where 
	$\EntitySet=\ItemSet\cup\PropSet$ is the set of entities 
	which can be subjects of a statement and $\ValueSet=\EntitySet\cup\DataValueSet$ 
	is the set of possible values of a property.
\end{definition}

In practice, Wikibase graphs also add the constraint that 
every item $q\in\ItemSet$ (or property $p\in\PropSet$) 
has a unique integer identifier $q^i\in\Nat$ ($p^i\in\Nat$).
In the Wikibase data model, statements contain a list of property-values and 
the values can themselves be nodes from the graph. 
This is different from property graphs, where the set of vertices and the set of values are disjoint. 

\begin{example}[Wikibase graph example] \label{example:WikibaseGraph}
We focus on a subset of Wikidata that models information about Tim Berners-lee (\timBl, Q80) and some of its awards. 
\timBl~is an \instanceOf (P31)~\Human (Q5). 
His \birthPlace (P19)~was \London (Q84), whose \country (P27)~is \UK (Q145) and his \birthDate (P569)~was 1955. 
He has as \employer (P108) the \CERN twice. 
The first one has \start{}~date (P580) 1980 and \pEnd{}~date (P582) 1980, and the second one between 1984 and 1994. 
He was \awarded~(P166) with the \emph{Princess of Asturias} (\PA) award together with (\togetherWith, P1706) Vinton Cerf (\vintCerf)~\footnote{
The award was really obtained by Tim Berners-Lee, Vinton Cerf, Robert Kahn and Lawrence Roberts, we included here only the first two for simplicity}, 
the award was given at the point in time (\pointTime, P585) 2002. The \country~of that award is \Spain and Vinton Cerf was born in New Haven (\newHaven):
	
\begin{tabular}{ccl}
\ItemSet &= \{ &  \timBl, \vintCerf, \London, \CERN, \UK, \Spain, \PA, \Human \} \\
\PropSet &= \{ &  \birthDate, \birthPlace, \country, \employer, \awarded, \\
& & \start, \pEnd, \pointTime, \togetherWith, \instanceOf \} \\
\DataValueSet &= \{ & 1984,1994,1980,1955\} \\
$\StmtSet$ &= \{ & (\timBl, \instanceOf, \Human, \{\}), \\ 
& & (\timBl, \birthDate, 1955, \{\}), \\ 
& & (\timBl, \birthPlace, \London, \{\}), \\
& & (\timBl, \employer, \CERN, \{ \start:1980, \pEnd:1980 \}), \\
& & (\timBl, \employer, \CERN, \{ \start:1984, \pEnd:1994 \}), \\
& & (\timBl, \awarded, \PA, \{\pointTime: 2002, \togetherWith:\vintCerf\}), \\
& & (\London, \country, \UK, \{\}), \\
& & (\vintCerf, \instanceOf, \Human, \{\}) \\
& & (\vintCerf, \birthPlace, \newHaven, \{\}) \\
& & (\CERN, \awarded, \PA, \{ \pointTime: 2013 \}) \\
& & (\PA, \country, \Spain, \{ \}) \} \\
\end{tabular}
	
Figure~\ref{figure:WikibaseGraphExample} presents a possible visualization of the example wikibase graph.
	
\begin{figure}
\centering
\includegraphics[width=\textwidth]{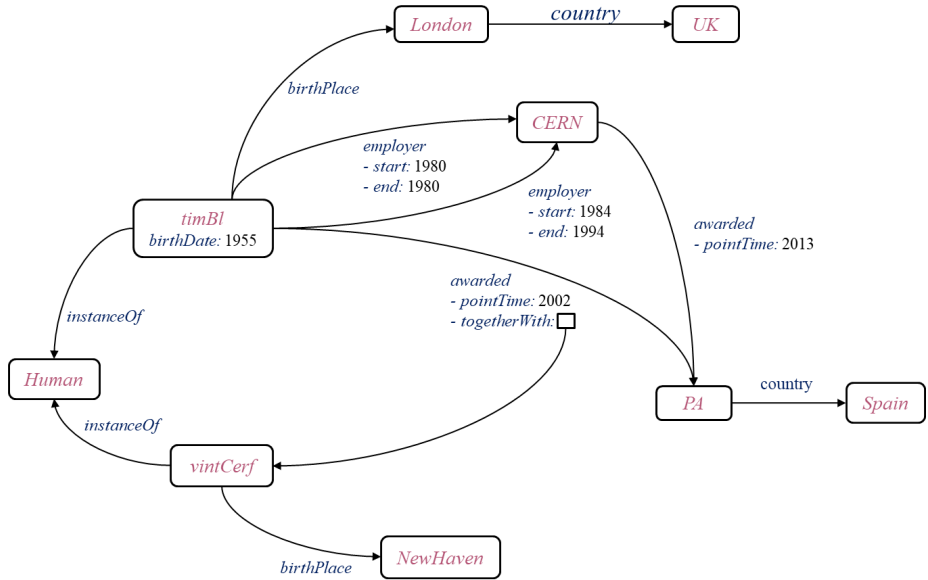}
\caption{Visualization of example wikibase graph} \label{figure:WikibaseGraphExample}
\end{figure}
	
\end{example}

The Wikibase data model supports 2 main export formats: JSON and RDF.
The JSON serialization directly follows the Wikibase data model. 
It basically consists of an array of entities 
where each entity is a JSON object that captures all the local information about the entity: 
the labels, descriptions, aliases, sitelinks and statements that have the entity as subject.
Each JSON object is represented in a single line.
A remarkable feature of this encoding is that it captures the output neighborhood of 
every entity in a single line making it amenable to processing models that focus on local neighborhoods.


%


The RDF serialization\footnote{\url{https://www.mediawiki.org/wiki/Wikibase/Indexing/RDF_Dump_Format}} was designed with the goal of being able to represent all the structures of the Wikibase data model in RDF, maintaining compatibility with semantic web vocabularies like RDFS and OWL and avoiding the use of blank nodes~\cite{EGKMV2014}. 

\begin{example}[RDF serialization of a node]
As an example, a fragment of the information about Tim Berners-Lee that declares that he is an instance of Human, has birth place London, has birth date 1955 and has employer with value CERN between 1984 and 1994 is represented in RDF (Turtle)~\footnote{The full Turtle serialization can be obtained at:~\url{https://www.wikidata.org/wiki/Special:EntityData/Q80.ttl}} as:
	
\begin{lstlisting}[language=Turtle]
wd:Q80 rdf:type wikibase:Item ;
 wdt:P31  wd:Q5   ;  # instance of = Human
 wdt:P19  wd:Q84  ;  # birthplace = London
 wdt:P569 "1955-06-08T00:00:00Z"^^xsd:dateTime ; # birthDate
 wdt:P108 wd:Q42944 ;  # employer = CERN
 p:P108 :Q80-4fe7940f   .
	
:Q80-4fe7940f rdf:type wikibase:Statement ;
 wikibase:rank wikibase:NormalRank ;
 ps:P108       wd:Q42944 ;
 pq:P580       "1984-01-01T00:00:00Z"^^xsd:dateTime ;
 pq:P582       "1994-01-01T00:00:00Z"^^xsd:dateTime .
\end{lstlisting}
	
The RDF serialization uses a direct arc to represent the preferred statement 
represented by prefix alias \lstinline|wdt:| 
leaving the rest of the values of a property accessible through the namespaces \lstinline|p:|, \lstinline|ps:| and \lstinline|pq:|.
The reification model employed by Wikidata creates auxiliary nodes that represent each statement. In the previous example, the node \lstinline|:Q80-4fe7940f| represents the statement which can be qualified with the start and end time. 
	
\end{example}

The RDF serialization model is employed in Wikidata to follow the linked data principles that enable to have a logical URI of a concept seprated from it's representation in different formats like HTML, JSON or RDF. 
It is also employed by the Wikidata Query Service which allow users to retrieve 
data using the SPARQL endpoint~\cite{BGK2018,MKGGB2018} and 
by RDF-based Wikidata dumps.

Wikidata adopted entity schemas using ShEx in a new namespace (schema entities start by letter \c|E| followed by a number). 
As an example, listing~\ref{EntitySchemaExample} presents an entity schema for researcher entities\footnote{The entity schema has also been created in Wikidata as:\href{https://www.wikidata.org/wiki/EntitySchema:E371}{E371}}.
Lines 1--6 contain prefix declarations following Turtle tradition. Lines 8--24 declare a \c|<Researcher>| shape which in this case has 7 triple constraints. 
The first constraint (line 9) states that items that conform to \c|<Researcher>| must be instances of Humans. 
The next line declares that the values of birth place (\c|wdt:P19|) must conform to shape \c|<Place>| declared in line 25.
Line 11 declares that the values of property \c|wdt:P569| must belong to datatype \c|xsd:dateTime|. The question mark indicates that they are optional.
Line 12 declares that the values of property \c|wdt:P108| (employer) must conform to shape \c|<Organization>| which is defined in line 28 (it is empty in this case). The star at the end indicates that there can be zero or more statements about \c|wdt:P108|.
Lines 13--17 declare the constraints on the qualifiers, in this case, that it is optional to have a \c|pq:P580| (start) time and a \c|pq:P582| end time statement. Notice that these declarations about qualifiers resemble the RDF serialization model which require to repeat the value of the property \c|p:P108| and \c|ps:P108| for the statement.  
Lines 18--23 follow a similar pattern.



%
%

\section{WShEx}

\begin{figure}
\centering
\includegraphics[width=0.7\textwidth]{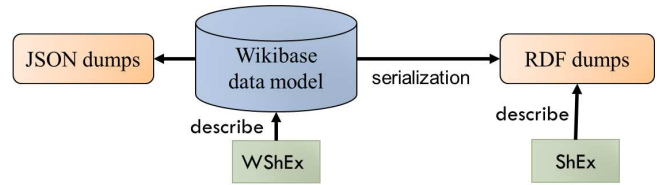}
\caption{Relationship between ShEx, WShEx and Wikibase data model} \label{fig:ShEx_vs_WShEx}
\end{figure}

As we have seen in the previous section, entity schemas in ShEx require users to be aware of how qualifiers and references are serialized in RDF which can lead to some duplication in their definition making their definitions more verbose than necessary. 
Another problem of ShEx schemas is that they cannot be used to directly describe the contents of Wikidata dumps in JSON which follow the Wikibase data model. 
In order to solve those issues, we propose an variant of ShEx called WShEx that can be used describe and validate the Wikibase data model. 
In this way, WShEx can also be used to validate Wikibase dumps in JSON without requiring them to be serialized in RDF.
Figure~\ref{fig:ShEx_vs_WShEx} represents the relationship between ShEx and WShEx.

In the following sections we present a simplified formal definition of WShEx adapted to wikibase graphs as defined in~\ref{def:wikibaseGraph} by presenting an abstract syntax followed by its semantic specification. 

\subsection{WShEx Abstract Syntax}

A \emph{WShEx Schema} is defined as a tuple $\langle\LabelSet,\schemaDef\rangle$ 
	where 
	$\LabelSet$ set of shape labels, 
	and $\schemaDef : \LabelSet\rightarrow\ShapeSet$ is a total function from labels to w-shape expressions where the set of shape expressions $se\in\ShapeSet$ is defined using the following abstract syntax:

\begin{tabular}{ccll}
$se$   & ::= &  cond  & Basic boolean condition on nodes (node constraint)\\
& $|$ & $s$ & Shape \\
& $|$ & $se_1$ \c|AND| $se_2$  & Conjunction \\
& $|$ & @\lbl & Shape label reference for $\lbl\in\LabelSet$ \\
 $s$    & ::= & \c|CLOSED| $s'$ & Closed shape \\
& $|$ & $s'$ & Open shape \\
		$s'$   & ::= & \{ te \} & Shape definition \\
		$te$ & ::= & \eachOf{te_1}{te_2} & Each of $te_1$ and $te_2$ \\
		& $|$ & \oneOf{te_1}{te_2} & Some of $te_1$ or $te_2$ \\
		& $|$ & $te*$ & Zero or more ${te}$  \\
		& $|$ & \arc{p}{@\lbl\,\,qs} & Triple constraint with predicate $p$ \\
		& & & value conforming to $\lbl$ and qualifier specifier $qs$ \\
		& $|$ & $\epsilon$  & Empty triple expression\\
		$qs$ & ::= & \openQs{ps}  & Open property specifier \\
		& $|$ & \closeQs{ps}      & Closed property specifier \\
		$ps$ & ::= & \eachOfQs{ps}{ps} & \emph{EachOf} property specifiers \\
		& $|$ & \oneOfQs{ps}{ps}  & \emph{OneOf} property specifiers \\
		& $|$ & ps* & zero of more property specifiers \\
		& $|$ & $\epsilon$ & Empty property specifier \\
		& $|$ & p:@\lbl & Property $p$ with value conforming to shape $\lbl$ \\
\end{tabular}

	
A ShEx schema that describes the Wikibase graph presented in example~\ref{example:WikibaseGraph} can be defined as:
	
\begin{tabular}{ccll}
\LabelSet & = & \{ & $Person$,$Place$,$Country$,$Organization$,$Date$, $Award$\} \\
\schemaDef($Person$) & = & \{ & \arc{birthDate}{@Date}; \arc{birthPlace}{@Place}; \\
 &   &  & \arc{employer}{@Organization} \openQs{\eachOfQs{start:@Date}{end:@Date}}* \\
 &   &  & \arc{awarded}{@Award} \openQs{\eachOfQs{pointTime:@Date}{togetherWith:@Person}}* \\
 & & \} & \\
\schemaDef($Place$)  & = & \{ & \arc{country}{@Country}\} \\
\schemaDef($Country$)  & = & \{ & \} \\
\schemaDef($Award$)  & = & \{ & \arc{country}{@Country}\} \\
\schemaDef($Organization$)  & = & \{ & \} \\
\schemaDef(Date)  & = &  & $\in{}xsd:date$\\
\end{tabular}

It is possible to define a compact syntax for WShEx in a similar way to ShExC adding the symbols $\lstinline!{|...|}!$ 
to declare open qualifier specifiers~\footnote{We adopted the same symbols as RDF-$\star$ (\url{https://w3c.github.io/rdf-star/cg-spec})} and 
$\lstinline![|...|]!$ for closed ones.

\lstset{language=WShEx}
Listing~\ref{WShExExample} presents a WShEx example that corresponds with the entity schema presented in listing~\ref{EntitySchemaExample}. 
It uses the default prefix declaration for items and properties. 
Lines 3--15 declare a \c|<Researcher>| shape with 5 triple constraints. 
The first one, declares that it must be instance of Human. 
The second one declares that they must have a \emph{birth place} (\c|:P19|) that conforms with shape \c|<Place>| declared in line 16. 
The third constraint indicates that the values of \emph{birth date} (\c|:P569|) must be \c|Time| values with a question mark indicating that it is optional. 
In WShEx it is possible to use Wikibase built-in datatypes like \c|Time|. Other values could be: \c|String|, \c|Item|, \c|Property|, \c|Quantity|, \c|MonolingualText|, \c|URL|, etc.\footnote{See~\url{https://www.wikidata.org/wiki/Help:Data_type} for a list of Wikibase built-in datatypes}. Lines 8--10 contain \emph{start} (\c|:P580|) and emph{end} (\c|:P582|) time qualifiers for the \emph{employer} (\c|:P108|) statement. 
Similarly, lines 12--14 declare \emph{point in time} (\c|:P585|) and \emph{together with} (\c|:P1706|) qualifiers for the \emph{award} statement. 

\lstset{basicstyle=\small}

\noindent\begin{minipage}{.48\textwidth}
\begin{lstlisting}[language=ShExC, caption=Example of ShEx-based Entity schema, label={EntitySchemaExample}, numbers=left, numberstyle=\tiny]
PREFIX pq: <.../prop/qualifier/>
PREFIX ps: <.../prop/statement/>
PREFIX p: <.../prop/>
PREFIX wdt: <.../prop/direct/>
PREFIX wd: <.../entity/>
PREFIX xsd: <...XMLSchema#>

<Researcher> {
 wdt:P31  [ wd:Q5 ]         ;
 wdt:P19  @<Place>          ;     
 wdt:P569 xsd:dateTime    ? ; 
 wdt:P108 @<Organization> * ;
 p:P108 { 
  ps:P108  @<Organization>   ;
  pq:P580  xsd:dateTime    ? ; 
  pq:P582  xsd:dateTime    ?
 } * ;
 wdt:P166    @<Award>   *     ; 
 p:P166 { 
  ps:P166  @<Award>         ;
  pq:P585  xsd:dateTime   ? ; 
  pq:P1706 @<Researcher>  *
 } *
}
<Place> { 
 wdt:P17 @<Country> 
}
<Organization>   {}
<Award> { 
  wdt:P17 @<Country> ?
}
<Country> {}
\end{lstlisting} 
\end{minipage}\hfill
\begin{minipage}{.48\textwidth}
\begin{lstlisting}[language=WShEx, caption=Example of WShEx schema, label={WShExExample}, numbers=left, numberstyle=\tiny]
PREFIX : <.../entity/>

<Researcher> {
 :P31     [ :Q5 ]            ;
 :P19     @<Place>          ;     
 :P569    Time     ? ; 
 :P108    @<Organization> 
  {| :P580 Time ? , 
     :P582 Time ?
  |} * ;
 :P166  @<Award>  
  {| :P585  Time   ? , 
     :P1706 @<Researcher> ?
  |} *
}
<Place> { 
 :P17 @<Country> 
}
<Organization> {}
<Award> { 
 :P17 @<Country> 
}
<Country> {}
\end{lstlisting}
\end{minipage}

\subsection{Semantics}

In order to define the semantics of WShEx 
we employ a conformance relation parameterized by a shape assignment $\conforms{\Graph}{n}{\typing}{se}$ with the meaning that node $n$ in graph $\Graph$ conforms to shape expression $se$ with shape assignment $\typing$ according to the rules~\ref{sem:WShEx_SE}.

\begin{table}[h!]
\begin{tabular}{c}
		\inference[$Cond$]
		{cond(n)=true} 
		{\conforms{\Graph}{n}{\typing}{cond}}
		
		\hspace{1cm}
		\inference[$AND$]
		{\conforms{\Graph}{n}{\typing}{se_1} & 
			\conforms{\Graph}{n}{\typing}{se_2}
		} 
		{\conforms{\Graph}{n}{\typing}{se_1 \text{ \texttt{AND} } se_2}}
		\\ \\
		
		%
		%
		
		\inference[$ClosedShape$]
		{ neighs(n,\Graph) = ts & 
			\conformsTE{\Graph}{ts}{\typing}{s'}
		} 
		{\conforms{\Graph}{n}{\typing}{$\c|CLOSED|\,$s'}}  
		\\ \\
		
		\inference[$OpenShape$]
		{ ts = \{\triple{x}{p}{y} \in neighs(n,\Graph) \mid p \in \props{te}\} &
			\conformsTE{\Graph}{ts}{\typing}{s'}
		} 
		{\conforms{\Graph}{n}{\typing}{s'}}
		\\ \\
	\end{tabular}
	\caption{Inference rules for WShEx shape expressions} \label{sem:WShEx_SE}
\end{table}

We also define a conformance relation $\conformsTE{\Graph}{ts}{\typing}{te}$ which declares that the triples $ts$ in graph $\Graph$ conform to the triple expression $te$ with the shape assignment $\typing$ using the rules~\ref{sem:WShEx_TEs} which takes into account qualifier specifiers. 

\begin{table}[h!]
	\begin{tabular}{c}
		\inference[$EachOf$]
		{ (ts_1,ts_2)\in\partition{ts} & 
			\conformsTE{\Graph}{ts_1}{\typing}{te_1} &
			\conformsTE{\Graph}{ts_2}{\typing}{te_2} 
		}
		{\conformsTE{\Graph}{ts}{\typing}{\eachOf{te_1}{te_2}}}
		
		\\ \\
		
		\inference[$OneOf_1$]
		{ \conformsTE{\Graph}{ts}{\typing}{te_1} }
		{\conformsTE{\Graph}{ts}{\typing}{\oneOf{te_1}{te_2}} }
		
		\hspace{1cm} 
		\inference[$OneOf_2$]
		{ \conformsTE{\Graph}{ts}{\typing}{te_2} }
		{\conformsTE{\Graph}{ts}{\typing}{\oneOf{te_1}{te_2}}}
		
		\\ \\
		\inference[$Star_1$]
		{}
		{\conformsTE{\Graph}{\emptyGraph}{\typing}{te*}}
		\\ \\
		\inference[$Star_2$]
		{(ts_1,ts_2)\in\partition{ts} & 
			\conformsTE{\Graph}{ts_1}{\typing}{te} &
			\conformsTE{\Graph}{ts_2}{\typing}{te*}}
		{\conformsTE{\Graph}{ts}{\typing}{te*}}
		
		\\ \\
		
		\inference[$TripleConstraint$]
		{ts = \{\quadruple{x}{p}{y}{s}\} & 
			\conforms{\Graph}{y}{\typing}{@\lbl} &
			\conformsQs{\Graph}{s}{\typing}{qs}
		}
		{\conformsTE{\Graph}{ts}{\typing}{\arc{p}{@\lbl}\,\,qs}}
		
	\end{tabular}
	\caption{Inference rules for WShEx triple expressions} \label{sem:WShEx_TEs}
\end{table}

Finally, the conformance relationship $\conformsQs{\Graph}{s}{\typing}{qs}$ 
between a graph $\Graph$ 
a set $s\in{}P\times{}V$ of property-value elements, 
a shape assignment $\typing$ and 
a qualifier specifier $qs$ is defined with the rules~\ref{sem:WShEx_Qs}. 

\begin{table}[h!]
	\begin{tabular}{c}
		\inference[$OpenQs$]
		{s'=\{(p,v)\in{}s|p\in{}preds(ps)\} &
			\conformsQs{\Graph}{s'}{\typing}{ps}
		}
		{\conformsQs{\Graph}{s}{\typing}{\openQs{ps}}} 
		
		\hspace{0.4cm}
		
		\inference[$CloseQs$]
		{\conformsQs{\Graph}{s}{\typing}{ps}}
		{\conformsQs{\Graph}{s}{\typing}{\closeQs{ps}}} 
		
		\\ \\
		
		\inference[$EachOfQs$]
		{\conformsQs{\Graph}{s}{\typing}{ps_1} &
			\conformsQs{\Graph}{s}{\typing}{ps_2}
		}
		{\conformsQs{\Graph}{s}{\typing}{\eachOfQs{ps_1}{ps_2}}} 
		
		\\ \\
		
		\inference[$OneOfQs_1$]
		{\conformsQs{\Graph}{s}{\typing}{ps_1}}
		{\conformsQs{\Graph}{s}{\typing}{\oneOfQs{ps_1}{ps_2}}} 
		
		\hspace{0.5cm}
		
		\inference[$OneOfQs_2$]
		{\conformsQs{\Graph}{s}{\typing}{ps_2} }
		{\conformsQs{\Graph}{s}{\typing}{\oneOfQs{ps_1}{ps_2}}} 
		
		\\ \\
		\hspace{-0.8cm}
		\inference[$StarQs_1$]
		{}
		{\conformsQs{\Graph}{\emptyset}{\typing}{ps*}} 
		
		\hspace{0.4cm}
		
		\inference[$StarQs_2$]
		{(s_1,s_2)\in{}\partition{s} & 
			\conformsQs{\Graph}{s_1}{\typing}{ps} &
			\conformsQs{\Graph}{s_2}{\typing}{ps*}
		}
		{\conformsQs{\Graph}{s}{\typing}{ps*}} 
		
		\\ \\
		
		\inference[$EmptyQs$]
		{}
		{\conformsQs{\Graph}{\emptyset}{\typing}{\epsilon}} 
		
		\hspace{0.5cm}
		
		\inference[$PropertyQs$]
		{ s = \{(p,v)\} &
			\conforms{\Graph}{v}{\typing}{@\lbl}
		}
		{\conformsQs{\Graph}{s}{\typing}{p:@\lbl}} 
		
	\end{tabular}
	\caption{Inference rules for WShEx qualifier expressions} \label{sem:WShEx_Qs}
\end{table}

\subsection{Implementation and use cases}

WShEx is currently implemented as a module\footnote{\url{https://github.com/weso/shex-s/tree/master/modules/wshex}} inside the 
ShEx-s project\footnote{\url{https://www.weso.es/shex-s/}}, a Scala implementation of ShEx.

The implementation includes a matcher for Entity documents which can be used to validate Wikidata dumps in JSON format. 
 In fact, the initial motivation for WShEx was the possibility to validate JSON dumps instead of RDF dumps to create Wikidata subsets~\cite{Labra2021KGSubsets}. 

One practical aspect that appeared is the need of a converter between Entity schemas defined in ShEx and WShEx. We have already implemented a first version of this converter.  
This approach allows to leverage the existing entity schemas which are defined in ShEx, convert them to WShEx and use them to process Wikibase JSON dumps. 

\section{Related work}

Our definition of Wikibase graphs was inspired by the formal definitions used for knowledge graphs in books like~\cite{Hogan2021}, which define two main data models: directed labeled graphs, which resemble RDF-based graphs and property graphs. 
We were also inspired by MARS (Multi-Attributed Relational Structures)~\cite{MKT2017}, which present a 
a generalized notion of property graphs adapted to Wikidata. 
In that paper, they also define MAPL (Multi-Attributed Predicate Logic) as a logical formalism that can be used for ontological reasoning.

Since the appearance of ShEx in 2014, there has been a lot of interest about RDF validation and description. 
In 2017, the data shapes working group proposed SHACL (Shapes Constraint Language) as a W3C recommendation~\cite{SHACLSpec}. 
Although SHACL can be used to describe RDF, its main purpose is to validate and check constraints about RDF data.
ShEx was adopted by Wikidata in 2019 to define entity schemas~\cite{ThorntonSSGMPW19}. We consider that ShEx adapts better to describe data models than SHACL, which is more focused on constraint violations. 
A comparison between both is provided in~\cite{Labra2017} while in~\cite{Labra-Gayo2019}, a simple language is defined 
that can be used as a common subset of both.

Improving quality of Knowledge graphs in general, and Wikidata in particular, has been the focus of some research like~\cite{Piscopo2019,Turki2020,Shenoy21}.



\section{Conclusions and future work}

WShEx can be used to describe and validate Wikibase graphs using an extension of Shape Expressions that handle also qualifiers.
We consider that WShEx schemas are more succint and adapt better to 
 the Wikibase data model. 
The language has been partially implemented and is being used to create Wikidata subsets from JSON dumps.
Future work is still necessary to 
 finish the implementation including a full grammar for the compact syntax, 
 more complete support for other Wikibase constructs like labels, descriptions, aliases, other built-in datatypes, ranks, etc. and 
 implement a full validator for Wikibase graphs based on WShEx. 

%

\printbibliography

\end{document}